\documentclass[twocolumn,floats,aps,epsfig,nofootinbib,amssymb]{revtex4}
\usepackage{latexsym,graphicx,epsfig,psfrag,float,amsmath,longtable,amssymb}
\usepackage{cancel,textcomp,bm,times,graphics,hyperref}
\pdfoutput=1
\allowdisplaybreaks
\usepackage{color}

\begin{document}


\title{\Large{$U(2)^5$ flavor symmetry and lepton universality violation in $W \rightarrow \tau \overline{\nu}_{\tau}$}}
\author{Alberto Filipuzzi and Jorge Portol\'es}
\affiliation{Departament de F\'{\i}sica Te\`orica, IFIC, CSIC --- Universitat de Val\`encia, Apt. Correus 22085, E-46071 Val\`encia, Spain}
\author{Mart\'in Gonz\'alez-Alonso}
\affiliation{Theoretical Nuclear, Particle, Astrophysics, and Cosmology (NPAC) Group, Department of Physics, University of Wisconsin-Madison, WI 53706, USA}
\date{\today}

\begin{abstract}
The seeming violation of universality in the $\tau$ lepton coupling to the W boson suggested by LEP II data is studied using an Effective Field Theory (EFT) approach.
Within this framework we explore how this feature fits into the current constraints from electroweak precision observables using different assumptions about the flavor structure of New Physics, namely $[U(2)\times U(1)]^5$ and $U(2)^5$. We show the importance of leptonic and semileptonic tau decay measurements, giving $3-4$ TeV bounds on the New Physics effective scale at $90\%$ C.L. We conclude under very general assumptions that it is not possible to accommodate this deviation from universality in the EFT framework, and thus such a signal could only be explained by the introduction of light degrees of freedom or New Physics strongly coupled at the electroweak scale.

\end{abstract}


\maketitle

\section{Introduction}

One of the features of the Standard Model (SM)  of particle physics is the universality of the lepton couplings, i.e. the fact that the coupling of the $W^{\pm}$ to the leptons does not depend on their flavor. However the experimental results from LEP-II on this issue \cite{Heister:2004wr,Achard:2004zw,Abdallah:2003zm,Abbiendi:2007rs,Alcaraz:2006mx} showed a slight deviation from universality coming from the third family, giving \cite{Nakamura:2010zzi}:
\begin{equation} \label{eq:univvio}
 R^W_{\tau\ell}=\frac{2 \, \mbox{BR} \left(W \rightarrow \tau \, \overline{\nu}_{\tau} \right)}{\mbox{BR}  \left( W \rightarrow e \, \overline{\nu}_{e} \right) + \mbox{BR} \left(W \rightarrow \mu \, \overline{\nu}_{\mu} \right)}
= 1.055 (23),
\end{equation}
resulting in $2.4$ standard deviations\footnote{The result given in Eq.~\eqref{eq:univvio} is obtained from the PDG fit to the branching ratios of the W \cite{Nakamura:2010zzi}, that uses LEP2 and $p\bar{p}$ colliders data. It is worth mentioning that considering only LEP2 data the discrepancy grows to 2.8 $\sigma$ \cite{Alcaraz:2006mx} (2.6 $\sigma$ using only published data \cite{LEPNote}),  all correlations included.} (all correlations included) from the SM prediction  $R^W_{\tau\ell}|_{\mbox{\tiny SM}}= 0.999$ \cite{Kniehl:2000rb}, which uncertainty is negligible compared with the experimental error. Recalling also the following ratio:
\begin{equation}
\label{eq:3br}
R^W_{\mu e} = \mbox{BR}\left( W \rightarrow \mu \, \overline{\nu}_{\mu} \right) / \mbox{BR}\left( W \rightarrow e \, \overline{\nu}_{e} \right) =   0.983 (18),
\end{equation}
and the correspondent SM prediction $R^W_{\mu e}|_{\mbox{\tiny SM}}= 1.000$, it can be concluded that the two lightest families seem to attach to the universality principle. The confirmation or refutation of this measurement is obviously very important, since such a violation by the third family would be a clear indication of New Physics (NP)  \cite{Pich:2008ni,Pich:2009zza}. However, it will not be easy for the LHC to reach such a precision in this observable, given the theoretical uncertainties associated to a hadronic machine. For this reason it is interesting to check indirectly this anomaly through its interplay with other related measurements. 
\par
Precision electroweak observables (EWPO), as well as other precise low energy measurements, provide constraints on new models looking for deviations that could foresee the NP structure. We study in this article if it is possible to accommodate the apparent discrepancy on the $W\to\tau\bar{\nu}_\tau$ channel within the present situation provided by EWPO, where essentially no disagreements have been found. In particular, lepton universality has been tested successfully at the per-mil level in $Z\to\ell^+ \ell^-$ \cite{Nakamura:2010zzi} and $\tau\to\nu_\tau\ell\nu_\ell$ decays (see e.g. Table 3 in Ref.~\cite{Pich:2012sx}), what makes very challenging to find a NP explanation for the large anomaly shown in Eq.~\eqref{eq:univvio}. Just for the sake of illustration, we show the values obtained in leptonic Z decays \cite{Nakamura:2010zzi}:
\begin{eqnarray}
\label{eq:Zll}
\mbox{BR}\!\left( Z \!\rightarrow \mu^+ \mu^- \right) \!/ \mbox{BR}\!\left( Z \!\rightarrow e^+ e^- \right) \!&=&\!   1.001 (3), \nonumber \\
\mbox{BR}\!\left( Z \!\rightarrow \tau^+ \tau^- \right) \!/ \mbox{BR}\!\left( Z \!\rightarrow \mu^+ \mu^- \right) \!&=&\!  1.001 (3),
\end{eqnarray}
in good agreement with the SM predictions, $1.000$ and $0.998$ respectively.
\par
Instead of adhering to a specific model we will follow an Effective Field Theory approach, where NP is parameterized by a tower of higher-dimensional operators~\cite{Leung:1984ni,Buchmuller:1985jz,Grzadkowski:2010es}. All NP theories which spectrum does not contain new light mass physical states (in comparison to those of the SM), that are weakly coupled at the electroweak scale and invariant under the SM gauge symmetries reduce at lower energies to the same effective Lagrangian, feature that makes this EFT approach very appealing.
Guided by the above-mentioned experimental data on lepton universality, we will consider different frameworks where the New Physics does not affect operators involving first and second generation fermions. As we will explain, this can be implemented through the adoption of specific flavor symmetries.
\par
For the numerical analysis, we greatly benefit from Ref.~\cite{Han:2005pr}, where constraints on these effective operators were obtained via a global fit to precision electroweak data. We modify the associated fitting code to introduce additional observables and operators\footnote{The code can be freely downloaded at the web page {\tt http://ific.uv.es/lhcpheno/}.}. These fit procedures are a powerful tool to analyze the impact of current constraints on different models.
\par
We apply this method to study the possible NP effects in leptonic W decays allowed by electroweak precision data. We will emphasize the role that the leptonic tau decay and the exclusive channel $\tau^-\to\pi^-\nu_\tau$ play in constraining specific directions of the parameter space of our theory, and the need to include these observables in this kind of analyses. We will see that the observed departure from universality cannot be accommodated within the current experimental scenario under quite general assumptions. Thus in order to be able to explain the observed deviation from lepton universality as a genuine NP effect, it seems to be necessary to resort to a different description of NP that could involve the introduction of new light degrees of freedom or a strongly interacting sector.
\par
A closely related issue driven by the $W \ell \nu_{\ell}$ vertex is the ratio of widths involving the leptonic decays of pseudoscalar heavy mesons,  $P \rightarrow \ell \, \overline{\nu}_{\ell}$. Accordingly, if any violation of universality is at work it also should be exposed in ratios of these decays into different charged leptons. Similarly, any modification of the SM coupling of $W$ with the tau lepton could show up, due to gauge symmetry, in the anomalous magnetic moment of the tau. We will comment how our results translate into these subjects.
\par
In the next section the EFT framework is introduced, along with different flavor symmetries and the relevant effective operators. In Sec.~\ref{sec:Wtaunu} we identify the operators that can generate a lepton universality violation in the third family, whereas in Sec.~\ref{sec:fit} we analyze through a global fit the bounds on these operators from EWPO and other low-energy measurements. Sec.~\ref{sec:heavymesons} is devoted to study the sensitivity of the leptonic decays of heavy mesons to the lepton universality violation, and Sec.~\ref{sec:conclusions} contains our conclusions. An Appendix collects several theoretical expressions not included in the main text.

\section{The Effective Field Theory framework}
\label{section:EFT}

Effective Field Theories embody the features, and particularly the dynamics, of the underlying theory. The astonishing performance of the SM suggests that whatever theory we find at higher energies has to reduce, upon integration of the relevant heavier degrees of freedom, to the key properties of the SM: symmetries and fields, that become its EFT. It is clear, though, that this approach breaks down if the underlying new physics contains physical states with mass $M \ll 1 \, \mbox{TeV}$, possibility that we do not consider in the present analysis. In this case the appropriate EFT should include that spectrum and its dynamics. In order to properly define our EFT setting we need moreover to assume that the new theory above the SM is weakly coupled at the weak scale, so that the gauge $SU(2)_L\times U(1)_Y$ symmetry is linearly realized.

\par The trail left in the procedure of integrating out heavier degrees of freedom is a Lagrangian with higher dimensional operators that respect its symmetry and content \cite{Leung:1984ni,Buchmuller:1985jz,Grzadkowski:2010es}:
\begin{equation} \label{eq:left}
 {\cal L}_{\mbox{\tiny EFT}} = {\cal L}_{\mbox{\tiny SM}} + \frac{1}{\Lambda} \sum_a \, \widehat{\alpha}_a^{(5)} \, {\cal O}_a^{(5)} \, +  \frac{1}{\Lambda^2} \sum_a \, \widehat{\alpha}_a^{(6)} \, {\cal O}_a^{(6)} ~+\ldots~,
\end{equation}
where ${\cal L}_{\mbox{\tiny SM}}$ is the SM Lagrangian, $\Lambda$ is the NP energy scale and ${\cal O}_a^{(n)}$ are $SU(2)_L\times U(1)_Y$ gauge-invariant operators of dimension $n$ built with SM fields (including the standard Higgs boson). Finally $\widehat{\alpha}_a^{(n)}$ are the dimensionless Wilson coefficients that carry the information of the underlying dynamics at the $\Lambda$ scale and are expected to be of ${\cal O}(1)$.
\par
The only gauge-invariant operator of dimension five violates lepton number, and thus it can be safely neglected under the assumption that the violation of that symmetry occurs at scales much higher than $\Lambda \sim 1$ TeV. Then the first order corrections to the SM predictions come from dimension-six operators. The contribution from these operators involve terms proportional to $v^2\slash\Lambda^2$, $v E \slash\Lambda^2$ and $E^2\slash\Lambda^2$, where $v \approx 174$ GeV is the vacuum expectation value of the Higgs field and E is the energy scale of the process considered. In order to be consistent with the truncation of the effective Lagrangian (\ref{eq:left}) we work at linear order in the above ratios, i.e. keeping only the contributions coming from the interference of the SM and dimension-six operators.
\par
In this article we consider the study of the apparent violation of universality in the couplings of W to leptons within the above EFT framework, with the goal of finding out if the observed deviation can be explained in terms of NP effects once constraints from precise electroweak observables are taken into account. Motivated by the data and for the sake of simplicity, we will assume two different flavor symmetries that we introduce in the next subsections.
\par
To set the stage for this discussion, we explain first the simpler case of $U(3)^5$ flavor symmetry.
In the absence of Yukawa couplings, the SM Lagrangian shows a $U(3)^5 = U(3)_{q}\times U(3)_{u}\times U(3)_{d}\times U(3)_{\ell}\times U(3)_{e}$ flavor symmetry,
corresponding to the independent rotation of each SM fermion field: the quark and lepton doublets $q$ and $\ell$ and the up-quark, down-quark and charged lepton
singlets $u$, $d$ and $e$. We can also decompose this symmetry group in the following way:
\begin{equation}
SU(3)^5\times U(1)_{L}\times U(1)_{B}\times U(1)_{Y}\times U(1)_{PQ}\times U(1)_{e}
\end{equation}
where the five global $U(1)$ symmetries can be identified with the total lepton and baryon number, the hypercharge, the Peccei-Quinn symmetry and a remaining global
 symmetry that we choose to be the rotation of the charged lepton singlet. In the presence of Yukawa couplings this flavor symmetry breaks down to the subgroup $G=U(1)_{L}\times U(1)_{B}\times U(1)_{Y}$.
\par
Requiring that the higher dimensional operators respect the $U(3)^5$ flavor symmetry reduces significantly their number, suppresses undesired
Flavor Changing Neutral Current effects and leads to the Minimal Flavor Violation (MFV) framework after the introduction of the Yukawa spurions \cite{D'Ambrosio2002ex}.
The complete list of the twenty-one dimension-six $U(3)^5$ invariant operators can be found in Refs.~\cite{Han:2004az,Cirigliano2009wk}, where this flavor symmetry was assumed in the context of an EFT analysis of electroweak precision data. As an example we show here the three operators that do not contain fermions:
\begin{eqnarray}
\label{eq:0f}
O_{W\!B}	&=&(h^\dagger \tau^a h) W^a_{\mu \nu} B^{\mu \nu}, \  \  \   O_h^3 = | h^\dagger D_\mu h|^2~, \nonumber \\
O_W 		&=& \epsilon_{abc} \, W^{a \nu}_{\mu} W^{b\lambda}_{\nu} W^{c \mu}_{\lambda}~,
\end{eqnarray}
where we follow, with minor modifications, the notation and conventions of Ref. \cite{Buchmuller:1985jz}:  $h$ is the Higgs boson doublet; $\tau^a$ are the Pauli matrices; $W_{\mu \nu}^i = \partial_{\mu} W_{\nu}^i - \partial_{\nu} W_{\mu}^i + g \, \varepsilon_{ijk} W_{\mu}^j W_{\nu}^k$, $B_{\mu \nu} = \partial_{\mu} B_{\nu} - \partial_{\nu} B_{\mu}$  and the covariant derivative reads $D_{\mu} = \partial_{\mu} -i \frac{g}{2} \tau^i W_{\mu}^i-ig' Y B_{\mu}$, with hypercharge $Y(h)=1/2$.
\par
It is clear that in this special framework it is impossible to generate any departure from lepton universality, as the $U(3)^5$ symmetry allows only for flavor independent NP contributions. For this reason we will relax this symmetry group to smaller groups where the third family is singled out.

\subsection{$[U(2)\times U(1)]^5$ flavor symmetry}
\label{sec:frameA}

Motivated by the experimental observations shown in Eqs.~\eqref{eq:univvio} and \eqref{eq:3br}, it is an interesting possibility to assume the flavor symmetry $[U(2)\times U(1)]^5$, that singularizes the third family with respect to the light ones, allowing for different NP contribution to the processes involving the heavy fermions: top, bottom and, in particular, $\tau$ and $\nu_\tau$.
\par
This framework was indeed studied in Ref.~\cite{Han:2005pr}, and we will use the same notation, in which $q_p$, $\ell_p$, $u_p$, $d_p$ and $e_p$ ($p=1,2$) represent only the two first generations of fermions, whereas $Q$, $L$, $t$, $b$ and $\tau$  represent the third family fields. The new notation makes clear which combinations of flavor indices are allowed by the flavor symmetry.  The operators that do not involve fermions are the same as in the $U(3)^5$ case, whereas those involving one or two fermion bilinears split in several operators; for instance:
\begin{eqnarray}
O_{he} = i (h^\dagger D^\mu h)(\overline{e} \gamma_\mu e) \rightarrow
&&\!\!\!\! \!  O_{he} = i (h^\dagger D^\mu h)(\overline{e} \gamma_\mu e) , \nonumber \\
&&\!\!\!\! \!  O_{h\tau} = i (h^\dagger D^\mu h)(\overline{\tau} \gamma_\mu \tau).
\end{eqnarray}
The list of invariant operators is much longer than in the $U(3)^5$ symmetric case, but not all the operators affect the EWPO. For this reason, and following Ref.~\cite{Han:2005pr},
we do not include in our numerical analyses (i) operators involving top quarks; (ii) operators involving only third-generation fermions; or (iii) operators involving light quarks and third generation leptons\footnote{We noticed that the operator $O_{Lq}^3$ in Eq.~(\ref{eq:hanope}) can be strongly constrained by the experimental value of the $\tau \to \pi \nu_{\tau}$ process and it is consequently included in our analysis. Ref.~\cite{Han:2005pr}, not considering this observable, didn't include $O_{Lq}^3$.}. Moreover, motivated by the experimental result shown in Eq.~\eqref{eq:3br} and for the sake of simplicity we will assume that 2- and 4-fermion operators that only have light generation fermions can be neglected. In this way we are left with the following six operators with one fermion bilinear:
\begin{eqnarray}
&&	O_{hf_1} = i (h^\dagger D^\mu h)(\overline{f}_1 \gamma_\mu f_1) + {\rm h.c.}~, \label{eq:12f} \\
&&	O_{hf_2}^{1} = i (h^\dagger D^\mu h)(\overline{f}_2 \gamma_\mu f_2) + {\rm h.c.}~,  \label{eq:22f} \\
&&	O_{hf_2}^{3} = i \left( h^{\dagger} D_{\mu} \tau^i h \right) \left( \overline{f}_2 \, \gamma^{\mu} \, \tau^i \, f_2 \right) + {\rm h.c.} ~,
\label{eq:2f}
\end{eqnarray}
where $f_1=\tau,b$ and $f_2=L,Q$. We also have the following four-fermion operators \cite{Han:2005pr}:
\begin{eqnarray} \label{eq:hanope}
  &&
  O_{Lq}^{3} = (\overline{L} \gamma^\mu \tau^i L) (\overline{q} \gamma_\mu \tau^i q) ~ ,
  O_{\ell Q}^{3}= (\overline{\ell} \gamma^\mu \tau^i  \ell) (\overline{Q} \gamma_\mu \tau^i  Q) ~,
  \nonumber\\
  &&
  O_{\ell Q}^{1}= (\overline{\ell} \gamma^\mu  \ell) (\overline{Q} \gamma_\mu  Q) ~, ~~~~~
  O_{Qe}=(\overline{Q} \gamma^\mu Q) (\overline{e} \gamma_\mu e) ~,
  \nonumber\\
  &&
  O_{eb}=(\overline{e} \gamma^\mu e) (\overline{b} \gamma_\mu b) ~,\qquad
  O_{\ell b}= (\overline{\ell} \gamma^\mu \ell) (\overline{b} \gamma_\mu b) ~,
  \nonumber\\
  &&
  O_{\ell L}^{1}= (\overline{\ell} \gamma^\mu \ell) (\overline{L} \gamma_\mu L) ~,~~~~~~
  O_{\ell L}^{3}= (\overline{\ell} \gamma^\mu \tau^i  \ell) (\overline{L} \gamma_\mu \tau^i  L) ~,
  \nonumber\\
  &&
  O_{Le}= (\overline{L} \gamma^\mu L) (\overline{e} \gamma_\mu e) ~, ~~~~~~
  O_{\ell\tau}= (\overline{\ell} \gamma^\mu \ell) (\overline{\tau} \gamma_\mu \tau) ~,
  \nonumber\\
  && O_{e\tau}=(\overline{e} \gamma^\mu e) (\overline{\tau} \gamma_\mu \tau) ~.
  \label{eq:4f}
\end{eqnarray}

\subsection{$U(2)^5$ flavor symmetry}

In the limit of vanishing Yukawa couplings the SM Lagrangian is invariant under the $[U(2)\times U(1)]^5$ group symmetry considered in the previous section. We can work with a more realistic scenario keeping the third family Yukawas $\mathcal{L} = y_t \bar{Q} \widetilde{h} t + y_b \bar{Q} h b + y_\tau \bar{L} h \tau$ and neglecting only those of the two lightest generations. In this case the flavor symmetry breaks down to $U(2)^5\times U(1)^3$, that we will just call $U(2)^5$, since the three $U(1)$ subgroups are simply the Lepton and Baryon number and Hypercharge of the third generation\footnote{A recent analysis of the implications of current flavor data for the quark-sector component of this symmetry, i.e. $U(2)^3$, suitably broken by spurions \`a la MFV, can be found in Ref.~\cite{Barbieri:2011ci}.}.

Among the new operators that appear due to the reduction of the symmetry group, only the following four chirality-flipping operators will affect EWPO:
\begin{eqnarray}
\label{eq:operwlnu}
O_{\tau B}^{t}  &=&  \left( \overline{L} \, \sigma^{\mu \nu} \, \tau \right) h \, B_{\mu \nu} + {\rm h.c.} ~, \nonumber \\
O_{b B}^{t}  &=&  \left( \overline{Q} \, \sigma^{\mu \nu} \, b \right) h \, B_{\mu \nu} + {\rm h.c.} ~, \nonumber \\
O_{\tau W}^{t}  &=&  \left( \overline{L} \, \sigma^{\mu \nu} \, \tau^i \, \tau \right) h \, W_{\mu \nu}^i + {\rm h.c.} ~, \nonumber \\
O_{b W}^{t}  &=&  \left( \overline{Q} \, \sigma^{\mu \nu} \, \tau^i \, b \right) h \, W_{\mu \nu}^i + {\rm h.c.} ~.
\end{eqnarray}
Their chirality-flipping structure translates, in the processes of our interest here, into contributions proportional to the fermion masses, i.e. suppressed by the factor $m_f / v$ with respect to other NP contributions from dimension-six operators. 
Given that we focus here on the $W \rightarrow \tau \, \overline{\nu}_{\tau}$ decay we will not consider in the following the operators $O_{bB}^{t}$ and $O_{bW}^{t}$.

\section{$W \rightarrow \tau \, \overline{\nu}_{\tau}$ decay in the EFT framework}
\label{sec:Wtaunu}

When the $U(2)^5$ flavor symmetry is assumed, the SM term and dimension-six operators contributing to the $W \rightarrow \tau \, \overline{\nu}_{\tau}$ decay are:
\begin{eqnarray}\label{eq:2ops}
 {\cal L}_{\mbox{\tiny EFT}} &=& i \, \overline{L}  D \! \! \! \!  /  L
 +\frac{1}{\Lambda^2} \left\{ \widehat{\alpha}_{hL}^{3} {\cal O}_{hL}^{3} + \widehat{\alpha}_{\tau W}^{t}{\cal O}_{\tau W}^{t} + \mbox{h.c.} \right\} \\ \hspace*{-0.7cm}
 \supset && \hspace*{-0.6cm} \frac{ g}{\sqrt2} \Big[ \Big(1 + 2 \alpha_{hL}^{3}  \Big) \bar{\tau}_L \gamma^\mu \nu_\tau W^-_\mu +  \frac{2}{g v}\alpha_{\tau W}^t \bar{\tau}_R \sigma^{\mu\nu} \nu_\tau W^-_{\mu\nu} \Big], \nonumber
\end{eqnarray}
where $W^-_\mu = (W^1_\mu + i W^2_\mu)\slash \sqrt2 $, 
and we have introduced the normalized couplings $\alpha \equiv \frac{v^2}{\Lambda^2} \,  \widehat{\alpha}$, that we assume to be real hereafter. Working at linear order in the $\alpha$ coefficients, the full decay width reads:
\begin{eqnarray} \label{eq:gammawlnu}
 \Gamma \left( W \rightarrow \tau \,\overline{\nu}_\tau \right) & = & \frac{G_F \, M_W^3}{6 \, \sqrt{2} \,  \pi} \left( 1- w_{\tau}^2 \right)^2  \\
 && \hspace{-2cm}\times \left\{ \left( 1 + 4 \alpha_{hL}^3 \right) \left(1+ \frac{w_{\tau}^2}{2} \right) + \, 6 \, \sqrt{2} \, w_{\tau} \, \alpha_{\tau W}^t \right. \biggr\} ~ , \nonumber
\end{eqnarray}
where $w_{\tau} = m_{\tau}/M_W$ and $G_F$ is the tree level Fermi coupling constant defined by $G_F \slash \sqrt2 = g^2\slash (8 M_W^2)$. The new contributions to the decay width have the following features:
\begin{itemize}
\item There are only two dimension-six operators contributing to this process: ${\cal O}_{hL}^{3}$ and ${\cal O}_{\tau W}^{t}$. This can be seen if the equations of motion are properly used to reduce the number of operators in the effective basis, as done in Ref.~\cite{Grzadkowski:2010es}, instead of using directly all the operators appearing in the original list of Ref.~\cite{Leung:1984ni,Buchmuller:1985jz}.
\item The lepton universality feature of the SM implies that $g_{\tau} = g$. The operator ${\cal O}_{hL}^3$ simply shifts the SM result in such a way that its effect can be encoded in the following redefinition:
\begin{eqnarray}
\label{eq:gtau}
 g_\tau \equiv g \left( 1 + \delta g_\tau \right) = g \left( 1 + 2~ \alpha_{hL}^3 \right)~.
\end{eqnarray}
This operator is allowed in the two flavor symmetries that we consider.
\item The magnetic operator ${\cal O}_{\tau W}^t$ provides a new structure not present in the SM \cite{Bernabeu:1994wh,Rizzo:1997we,GonzalezSprinberg:2000mk}. Contrarily to ${\cal O}_{hL}^3$ this is a chirality flipping operator and it gives a contribution suppressed by $m_{\ell}/M_W$ due to the derivative dependence. Assuming the $[U(2)\times U(1)]^5$ flavor symmetry this term vanishes.
\end{itemize}
In what follows we will consider the universality ratios $R^W_{\ell\ell^\prime} = \Gamma(W\to\ell\nu_\ell) \slash \Gamma(W\to\ell^\prime\nu_\ell^\prime)$, instead of the simple decay rate, 
in such a way that we do not have to worry about the NP corrections associated to the experimental determination of the Fermi constant $G_F$, since they cancel in the ratio.

\section{Fit procedure and results}
\label{sec:fit}

Once we have identified in the previous section the effective operators that can contribute to $R_{\tau\ell}^W$, generating a deviation from lepton universality, we study now the constraints that can be derived on these operators from EWPO and low-energy measurements.
\par
Looking for example at the experimental result \eqref{eq:Zll} it can be understood that one single operator will not be able to explain simultaneously the EWPO and the anomaly in the $W\tau\nu$ vertex shown in Eq.~\eqref{eq:univvio}, due to the gauge symmetry that connects W and Z bosons. However, when several operators are present one can have cancellations between them and a careful numerical analysis is needed.
\par
With that purpose we updated and modified the Mathematica code developed in Ref.~\cite{Han:2005pr}, that included electroweak observables at the Z line and at higher energies and other low energy measurements. In addition we include the leptonic tau decay and the exclusive channel $\tau \to \pi \nu_\tau$, that have an experimental error well below the 1\% level and a theoretical error under control. We consider also the anomalous magnetic moment of the tau lepton that, despite its very large experimental uncertainty, is able to constrain the magnetic operators poorly bounded by other observables.  The associated formulas are collected in Appendix \ref{app:A} and the complete list of the observables used in our analysis can be found in Table \ref{tab:experiments}.
\begin{table}[tb]
\begin{tabular}{|l|l|l|l|}
\hline
 Classification 		& Std. Notation					&  Measurement \\
 \hline
 Atomic parity 		&$Q_W(Cs)$						&Weak charge in Cs\\
 violation 			& $Q_W(Tl)$						& Weak charge in Tl\\
 \hline
   DIS				&$g_L^2,g_R^2$					&$\nu_\mu$-nucleon scattering (NuTeV)\\
      				&$R^\nu$							&$\nu_\mu$-nucleon scatt. (CDHS, CHARM)\\
      				&$\kappa$							&$\nu_\mu$-nucleon scatt. (CCFR)\\
      				&$g_V^{\nu e},g_A^{\nu e}$		&$\nu$-$e$ scatt. (CHARM II)\\

 \hline
   Z-pole			&$\Gamma_Z$					&Total $Z$ width\\
				    	&$\sigma_0$ 						&$e^+e^-$ hadronic cross section\\
					&$R_{f=e,\mu,\tau,b,c}^0$		&Ratios of decay rates \\
        				&$A_{FB}^{0,f=e,\mu,\tau,b,c}$	&FB asymmetries\\
     				&$A_{f=e,\mu,\tau,s,b,c}$			&Polarized asymmetries\\
					&$\sin^2\theta_{eff}^{lept}$		&Hadronic charge asymmetry\\
 \hline
 LEPII				&$\sigma_{f=q,c,b\mu,\tau}$		& Total cross sections for $e^+e^-\!\rightarrow\! f\overline f$\\
 fermion			&$A_{FB}^{f=c,b,\mu,\tau}$		& FB asymmetries for $e^+e^-\rightarrow f\overline f$\\
 production		&$d\sigma_e/d\cos\theta$		&$e^+e^-\rightarrow e^+e^-$ diff. cross section\\
 \hline
 W pair				&$d\sigma_W/d\cos\theta$		&$e^+e^-\!\rightarrow\! W^+W^-$ diff. cross section\\
	 				&$M_W$							&W mass \\
\hline
$V_{\mbox{\tiny{CKM}}}$ unitarity		& $\Delta_{CKM}$							&  $V_{ud}$ and $V_{us}$ extractions \cite{Cirigliano2009wk}\\
\hline
$\tau$ decays		& $\tau\to \nu_\tau \ell \bar{\nu}_\ell$		&  Leptonic $\tau$ decay ($\ell = e, \mu$) \cite{Nakamura:2010zzi}\\
				    	& $\tau\to \nu_\tau \pi$						&  Exclusive hadronic $\tau$ decay \cite{Nakamura:2010zzi}  \\
\hline
Anomalous  		&   $a_\tau$									&   $e^+e^- \to e^+e^- \tau^+ \tau^-$ \\
magnetic 			&												&	cross section \cite{Abdallah:2003xd}\\
moment			&												&	\\		
\hline
\end{tabular}
\caption{\small{Measurements included in this analysis. See Ref.~\cite{Han:2004az} and references therein for detailed descriptions. References are shown only for the new observables.
\label{tab:experiments}}}
\end{table}
\par
We included in the program also the contribution to the different observables coming from the magnetic operators $O_{\tau W}^{t}$ and $O_{\tau B}^{t}$, not included in Ref.~\cite{Han:2005pr} since the $U(2)^5\times U(1)^5$ symmetry was assumed in that work. The formulas for the Z decay rate can be found in Appendix \ref{app:b}, whereas the formulas for $e^+ e^- \to \tau^+ \tau^-$ cross section have been taken from Ref.~\cite{GonzalezSprinberg:2000mk}.
\par
The leptonic decays of heavy pseudoscalar mesons ($B^\pm$, $D^\pm$, $D^\pm_{\mbox{\tiny S}}$) could in principle be considered in order to constrain NP effects in leptonic W decays, but they have not reached yet the necessary experimental precision: the relative error of the current data on the decays into tau are approximately $\mathcal{O}(6\%)$ for $D_{\mbox{\tiny S}}$ decays and $\mathcal{O}(20\%)$ for B decays, and some of the decays into muon and electron have not been seen yet, preventing a complete analysis of the lepton universality ratios. For these reasons, these observables have not been included in the fit. We will comment on them in Sec.~\ref{sec:heavymesons}.
\par
Concerning LHC measurements, the natural channels to analyze for the purpose of this paper are $pp\to \tau \bar{\nu} X$ and $pp\to \tau^+ \tau^-X$, where possible modifications of the $W\tau\nu$ vertex and its gauge counterpart $Z\tau^+\tau^-$ can be probed, but unfortunately there is no data available for these particular channels yet. On the theoretical side, the contribution to these processes coming from effective operators has been worked out in Refs.~\cite{Bhattacharya:2011qm,Cirigliano:2012} for first generation leptons. In any case, as we will see, the list of observables included in our fit is exhaustive enough to reach a solid answer to the possible lepton universality violations.
\par
With the above-mentioned observables $O^i$, we build a standard $\chi^2$ function as:
\begin{equation} \label{eq:chi2def}
\chi^2 \left(\boldsymbol{\alpha} \right) = \sum_i  \left[ O_{\mbox{\tiny th}}^i \left(\boldsymbol{\alpha} \right) - O_{\mbox{\tiny exp}}^i \right]
\left[\sigma_{\mbox{\tiny O}}^2
\right]^{-1}_{ij} \left[ O_{\mbox{\tiny th}}^j \left(\boldsymbol{\alpha} \right) - O_{\mbox{\tiny exp}}^j \right]   \, ,
\end{equation}
where the error matrix $\sigma^2_{\mbox{\tiny O}}$ includes the experimental error and the uncertainty on the SM prediction combined in quadrature. The theoretical value $O_{\mbox{\tiny th}}^i$ contains the up-to-date SM prediction and the contribution of higher dimensional operators through interference with SM vertices, i.e. linear in the  $\alpha_a$ couplings.
\par
As a result of this fit we determine the value of the different Wilson coefficients $\alpha_a$, with their relative errors and the corresponding correlations, or in other words the bounds on the different NP effective operators. In particular we are interested in the bounds associated to the two operators that could generate a lepton universality violation in the W decay (see Eq.~\eqref{eq:2ops}), and finally in the determination of the universality ratio $R_{\tau \ell}^W$ extracted from our fit, to be compared with the experimental determination given in Eq. (\ref{eq:univvio}).

\subsection{(Semi)leptonic $\tau$ decays as precise electroweak observables}

\begin{figure}[t]
\begin{center}
\includegraphics[width=7.7cm]{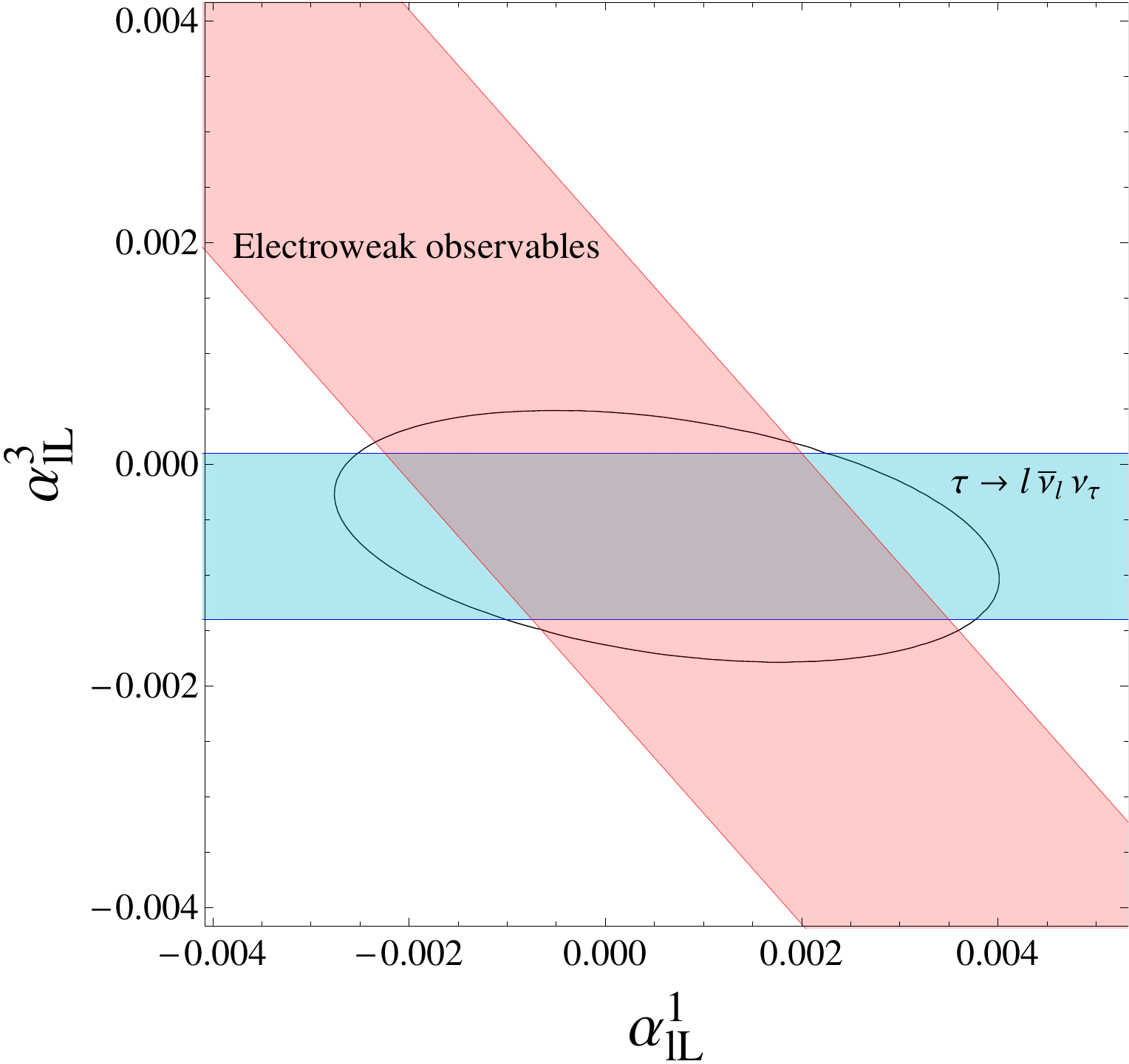}
\caption{\label{fig:TauConstrains}
Phenomenological constraints on the operators $O_{\ell L}^1$ and $O_{\ell L}^3$ from electroweak observables (red diagonal band) and leptonic tau decay rate (blu horizontal band). See Table \ref{tab:experiments} for the complete list of observables considered. The black ellipse is the 1$\sigma$ C.L. region when considering all observables together. }
\end{center}
\end{figure}

In a general analysis involving a big number of operators (free parameters in the fit) it is possible to encounter flat directions, i.e. directions in the parameter space that are not bounded by the experimental data. This means that some operators appear always in the same combination throughout all the observables considered in the fit and then only that combination can be constrained, and not each operator separately. In Ref.~\cite{Han:2005pr} four flat directions were identified in the particular fit we are using in this work. However, we show now how the addition of the leptonic tau decay to the list of EWPO included in the fit removes one of these flat directions.
\par
In the limit of $U(2)^5$ flavor symmetry the rate for the leptonic decay of the $\tau$ lepton reads\footnote{The operator corresponding to $\alpha_{h\ell}^3$ is defined in analogy with $O_{hL}^3$ in Eq.~\eqref{eq:2f}. This self-explanatory notation will be adopted hereafter for operators involving only light fermions. Although we neglect these operators in the subsequent numerical analysis we keep them in the analytic expressions for the sake of completeness.}:
\begin{eqnarray} \label{eq:taulepto}
\Gamma_{\tau\to\nu_\tau\ell\overline{\nu}_\ell} &=& \frac{G_F^2 m_\tau^5}{192 \, \pi^3}    \left\{ \Big[1+ 4 \, \alpha_{h L}^{3} + 4 \, \alpha_{h \ell}^{3} - 4 \, \alpha_{\ell L}^{3} \Big] \times \right.\nonumber\\
&& \hspace{-1.8cm} \times f\!\left(\frac{m_\ell^2}{m_\tau^2}\right)  \left. + 2 \, \sqrt{2} \, \alpha_{\tau W}^{t} \, \frac{m_\tau}{M_W} \, g\!\left(\frac{m_\ell^2}{m_\tau^2}\right) \right\} (1\!+\!\delta_{RC})~,
\end{eqnarray}
where $\ell=e,\mu$, $\delta_{RC}$ contains the radiative corrections to the SM contribution \cite{Pak:2008qt} and
\begin{eqnarray}
f(x) \!&=&\! 1 - 8x - 12 x^2 \ln{x} + 8x^3 - x^4, \\
g(x) \!&=&\! 1 - 6 x + 18 x^2 - 10 x^3 + 12 x^3 \ln{x} - 3 x^4 . \nonumber
\end{eqnarray}
\par
In order to show the constraining power of the tau decays let us consider the simple situation in which only the operators $O_{\ell L}^1$ and $O_{\ell L}^3$ are not vanishing. As shown in Fig.~\ref{fig:TauConstrains} the electroweak observables, and in particular the $e^+ e^- \to \tau^+ \tau^-$ cross section and the forward-backward asymmetry, are able to constrain only the combination $O_{\ell L}^1 + O_{\ell L}^3$. The inclusion of the the leptonic tau decay into the fit allows to reduce the one sigma C.L. region to the black ellipse. The two operators are then constrained at the $0.4\%$ and $0.2\%$ level, corresponding to an effective NP scale $\Lambda>2.7$ TeV and $\Lambda>4.1$ TeV (90$\%$ C.L.) respectively: very strong bounds that show the importance of leptonic tau decays as electroweak precision observable. 

A similar role is played by the pionic $\tau$ decay, where experimental results and SM calculations are also below the per-mil level of precision. The expression for the $\tau\to\pi\nu_\tau$ decay rate within our $U(2)^5$ flavor symmetric EFT framework is the following:
\begin{eqnarray}
\Gamma_{\tau^- \to \pi^- \nu_{\tau}} &&= \frac{G_F^2 F_{\pi}^2}{8\pi} |V_{ud}|^2 m_\tau^3 \left( 1-\frac{M_\pi^2}{m_\tau^2}\right)^2 (1+\delta_{RC}') \nonumber\\
&& \times \left( 1 + 4~\alpha_{hL}^{3} + 4~\alpha_{hq}^{3} - 4~\alpha_{L q}^{3}  \right)   ~,
\end{eqnarray}
where $F_\pi$ denotes the pion decay constant and $\delta_{RC}'$ radiative corrections \cite{Guo:2010dv}. It is convenient to work once again with a normalized ratio, namely:
\begin{eqnarray} \label{eq:rtaupio}
R_{\tau/\pi}
\label{eq:Rtaupi}
&\equiv& \frac{\Gamma_{\tau^- \to \pi^- \nu_{\tau}}}{\Gamma_{\pi^- \to \mu^- \overline{\nu}_{\mu}}} \nonumber\\
&=& \frac{m_\tau^3}{2 \, m_\mu^2 M_\pi} \left( \frac{1- \frac{M_\pi^2}{m_\tau^2}}{ 1- \frac{m_\mu^2}{M_\pi^2}}\right)^2 \left( 1 + \delta_{\tau/\pi} \right) \nonumber\\
&& \hspace{-0.5cm} \times  \left( 1 + 4\left(\alpha_{hL}^{3} - \alpha_{h\ell}^{3} \right) - 4\left(\alpha_{L q}^{3} - \alpha_{\ell q}^{3} \right)  \right)   .
\end{eqnarray}
where $\delta_{\tau/\pi}=0.0016(14)$ denotes the radiative corrections to the SM contributions \cite{Decker:1994dd}. As we can see, this observable represents another probe of the $\alpha_{hL}^3$ coefficient, and moreover it represents the only observable in our analysis sensitive to the $\alpha_{L q}^{3}$ coefficient. Comparing the experimental value of $R_{\tau/\pi}$ \cite{Nakamura:2010zzi,Schael:2005am} and its SM prediction we get a bound of $\Lambda > 3.1$ TeV (90$\%$ C. L.) on the NP effective scale for the four Wilson coefficients appearing in Eq.~\eqref{eq:Rtaupi}.

\subsection{$[U(2)\times U(1)]^5$ symmetric case: results}

\begin{figure}[t]
\begin{center}
\includegraphics[width=8cm]{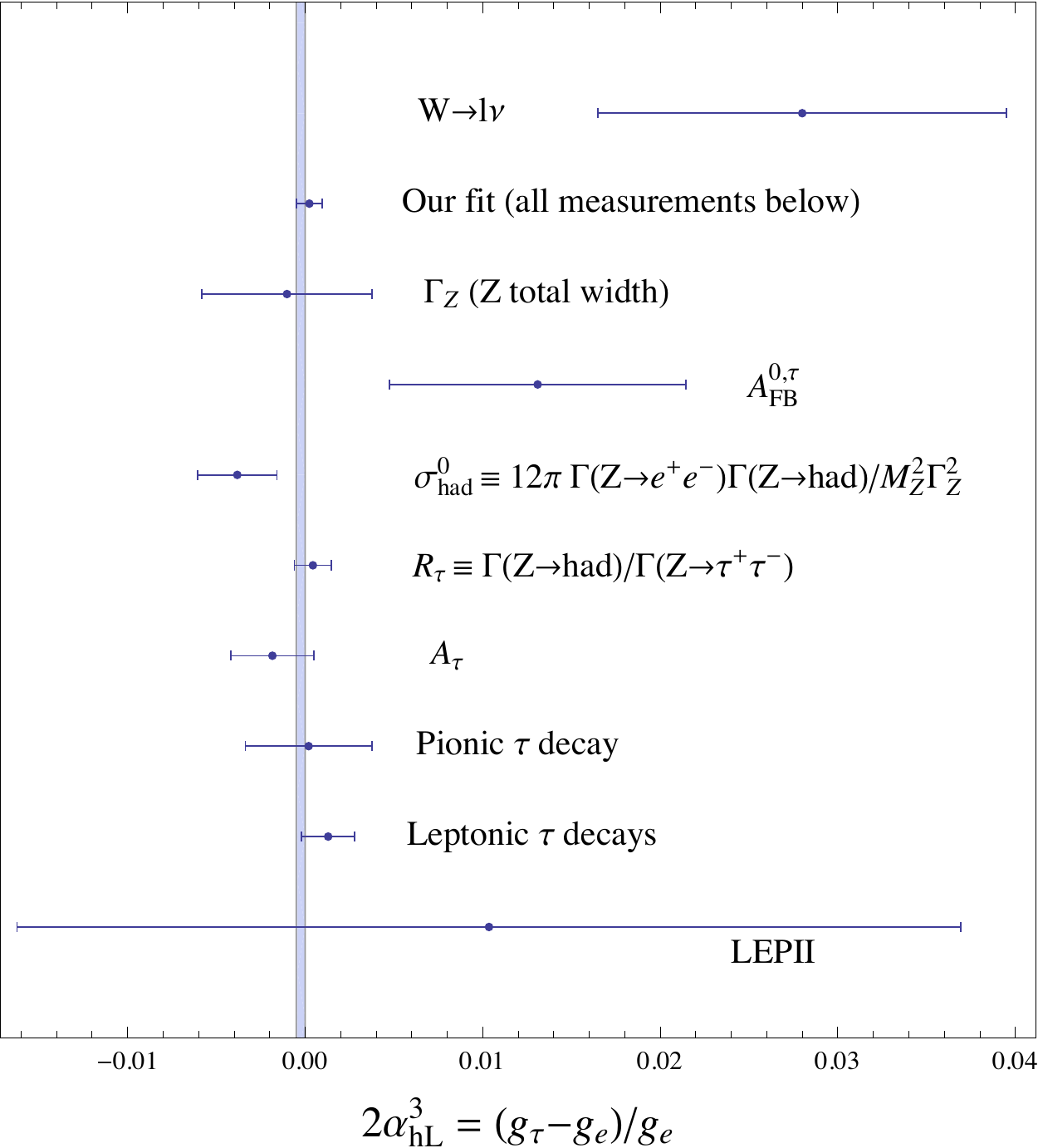}
\caption{\label{fig:SingleOperator}
Bounds obtained for the NP coefficient $\alpha_{hL}^3$ from the different set of measurements included in our fit. Equivalently, these are the bounds on the deviation from lepton universality in the electroweak coupling to third generation leptons $g_\tau$ (see Eq.~\eqref{eq:gtau}). For comparison we also show the value obtained, using the experimental data in Eq. (\ref{eq:univvio}), from leptonic W decays (not included in our fit). $A_{FB}^{0,\tau}$ is the forward-backward asymmetry measured at LEP1 for tau pairs, $A_\tau$ includes the SLD measurement and the LEP1 total $\tau$ polarization and the LEP2 bound comes from $\tau$ pair cross sections and asymmetries. See PDG \cite{Nakamura:2010zzi}, chapter 10, for more details.}
\end{center}
\end{figure}

In order to study if the $R_{\tau\ell}^W$ anomaly of Eq.~\eqref{eq:univvio} can be accommodated in our EFT framework as a genuine New Physics effect and not just a statistical fluctuation,  we start with a single operator analysis where only the $\alpha_{hL}^3$ is present and all the observables of Table \ref{tab:experiments} are included. In this case we obtain the expected strong bound:
\begin{equation}\label{eq:Rtau}
R_{\tau\ell}^W = 0.9997 \pm 0.0015~,
\end{equation}
in good agreement with the SM prediction. As shown in Fig.~\ref{fig:SingleOperator}, the very precise measurements of leptonic $Z$ and $\tau$ decays dominate our fit, and makes impossible to accommodate the $R_{\tau\ell}^W$ anomaly.

Once we include additional operators, things become less intuitive because cancelations between operators are possible, opening the possibility to explain the $R_{\tau\ell}^W$ anomaly and the leptonic Z and $\tau$ decays at the same time.

As a first global analysis, we assume the $[U(2)\times U(1)]^5$ flavor symmetry and we include the 17 operators given in Eqs.~\eqref{eq:0f} and (\ref{eq:12f}-\ref{eq:4f}). It is worth repeating that in order to simplify the discussion and given that the experimental data show no sign of NP related to the light families of fermions, we have assumed that the operators involving only light fermions can be neglected.
\par
Somehow surprisingly we find that even with so many operators, the constraint on $\alpha_{hL}^3$ is very strong, namely $- 3.6 \times 10^{-3} \leq \alpha_{hL}^{3} \leq - 0.5 \times 10^{-3} $ at $90\%$ C.L. Interestingly enough, this value is two sigmas away from zero, giving the following bound on the universality ratio:
\begin{equation}\label{eq:RtauBIS}
R_{\tau\ell}^W = 0.991 \pm 0.004 ~,
\end{equation}
where we quoted the error at 1$\sigma$ level in order to be comparable with the experimental result in Eq. (\ref{eq:univvio}). Thus we find the curious result that our fit is indeed able to accommodate a violation of lepton universality in the W decays, but \emph{in the opposite direction} than the direct experimental measurement. Somehow the small tensions present in the SM fit (see e.g. $\sigma_{had}^0$ in Fig.~\ref{fig:SingleOperator}) can be alleviated introducing some non-zero Wilson coefficients, being $\alpha_{hL}^3$ one of them. Obviously the inclusion of $R_{\tau\ell}^W$ will reduce this \lq\lq tension\rq\rq~moving the value of $\alpha_{hL}^3$ closer to zero ($- 3.2 \times 10^{-3} \leq \alpha_{hL}^{3} \leq - 0.08 \times 10^{-3} $ at $90\%$ C.L.).
\par
The conclusion is once more that we cannot accommodate the $R_{\tau\ell}^W$ along with our long list of precision observables, and thus we are forced to consider it a mere statistical fluctuation. 
Unlike the single operator case where it could be naively expected, this represents a non-trivial result in a fit with seventeen free parameters.
\par
For the sake of completeness, let us mention that in a truly global $[U(2)\times U(1)]^5$ fit, where operators only involving light fermions (like e.g. ${\cal O}_{h\ell}^3$) are also included, the NP bounds become extremely weak and the current experimental value of $R_{\tau\ell}^W$ cannot be excluded anymore.

\subsection{$U(2)^5$ symmetric case: results}

Reducing the symmetry group to $U(2)^5$ introduces the chirality-flipping operators $O_{\tau W}^t$ and $O_{\tau B}^t$, offering additional NP contributions to the observables and higher cancellations between operators.
\par
From the associated global fit\footnote{We do not include in this $U(2)^5$-symmetry fit the leptonic polarization asymmetries $A_\ell$, since they have been extracted assuming only vector and axial-vector couplings.} with 19 free parameters, we get the following $90\%$ bounds on the two operators involved in the W decays: $- 3.7 \times 10^{-3} \leq \alpha_{hL}^3 \leq - 0.6 \times 10^{-3} $ and $0.04 \times 10^{-3} \leq \alpha_{\tau W}^{t} m_\tau / M_W \leq 5.0 \times 10^{-3}$, where we have explicitly shown the $m_\tau \slash M_W$ suppression that multiplies the $\alpha_{\tau W}^t$ coefficient in the observables. From these values we calculate the prediction for the universality ratio at 1$\sigma$ level:
\begin{equation}
R_{\tau\ell}^W = 1.01 \pm 0.01 ~.
\end{equation}
While the constraints on $\alpha_{hL}^3$ are very similar to the previous case, the presence of a second contribution from the magnetic operator increases the error (and the central value) of $R_{\tau\ell}^W$. This increase is however not enough to nicely accommodate the experimental value of $R_{\tau\ell}^W$ shown in Eq.~\eqref{eq:univvio}.

\section{Leptonic decays of heavy mesons}
\label{sec:heavymesons}

The leading SM contribution to the $P^- \rightarrow \ell^- \, \overline{\nu}_{\ell}$ decays is given by the W exchange and hence it is interesting to point out how these decays get modified by possible deviations from family universality in the $W\ell\nu$ coupling. In particular we are interested in the $D$, $D_{\mbox{\tiny S}}$ and $B$ decays because they are heavy enough to decay into the tau lepton. Although two dimension-six operators modify the vertex of the $W$ gauge boson with leptons, namely ${\cal O}_{h \ell}^3$ and ${\cal O}_{e W}^t$, only the first contributes to the leptonic decay of heavy mesons, due to the fact that the tensor coupling has no spin-0 component.
\par
In order to get rid of the hadronic uncertainties and the NP corrections to the Fermi constant or the CKM elements appearing in the individual decay widths, we will focus again on the ratio between the tau channel and a light lepton channel:
\begin{eqnarray}
R_{\tau \ell}^P
&=& \frac{\mbox{BR} \left( P \rightarrow \tau \, \overline{\nu}_{\tau} \right)}{\mbox{BR} \left( P \rightarrow \ell \, \overline{\nu}_{\ell} \right)}~,
\end{eqnarray}
where $\ell=e,\mu$. The effective Lagrangian that mediates these decays, including linear corrections in the $\alpha$ coefficients, can be found in Eq. (34) of \cite{Cirigliano2009wk}. Assuming the $U(2)^5$ flavor symmetry we find the following expressions for the ratios\footnote{In the $B$ decays we neglect the contribution from a new $U(2)^5$-invariant operator $O_{Qb\tau} = (\bar{L} \tau) (\bar{b}Q) + \mbox{h.c.}$, since this operator does not affect the EWPO included in our fit.}:
\begin{eqnarray} \label{eq:pratio}
R_{\tau \ell}^{D_{(s)}}
\!\!\!&=&\!\! \frac{h_{D_{(s)}}\!\!\left(m_{\tau}\right)}{h_{D_{(s)}}\!\!\left(m_{\ell}\right)} \left\{ 1 + 4 \left(  \alpha_{hL}^3 \!-\! \alpha_{h\ell}^3 \right) - 4 \left(  \alpha_{Lq}^3 \!-\! \alpha_{\ell q}^3 \right) \right\} ~,\nonumber\\
R_{\tau \ell}^{B}
\!\!&=&\!\! \frac{h_{B}\!\left(m_{\tau}\right)}{h_{B}\!\left(m_{\ell}\right)} \left\{ 1 + 4 \left(  \alpha_{hL}^3 - \alpha_{h\ell}^3 \right) - 4 \left(  \alpha_{LQ}^3 - \alpha_{\ell Q}^3 \right) \right\} ~,\nonumber \\
\end{eqnarray}
where $h_P\left(m\right) = m^2 \left( 1-m^2/M_{P}^2 \right)^2$. 
\par
As expected, we find that the $\alpha_{hL}^3$ coefficient modifies these ratios. However, the bound on this coefficient from our analysis of EWPO and low-energy measurements is below the per-cent level (see Sec.~\ref{sec:fit}), a precision very far from current experimental results in these decays. The only ratio where we actually have a value, and not just an upper or lower limit, is $R_{\tau\mu}^{D_s} = 9.2(7)$ \cite{Nakamura:2010zzi}. We can compare this $\sim 8\%$ experimental error with the $\sim 0.7\%$ determination of the $R_{\tau/\pi}$ ratio, where exactly the same linear combination of NP couplings is probed, as shown in Eq.~\eqref{eq:Rtaupi}. This level of precision can actually be considered a benchmark sensitivity for future $D$ and $D_s$ meson experiments to become competitive in the NP search within our EFT framework.

On the other hand the leptonic $B$ decays are interesting since they probe a different linear combination of NP coefficients, and therefore are complementary to other observables.

\section{Conclusions}
\label{sec:conclusions}

In the SM the coupling of leptons to the gauge bosons is flavor blind, a property that has been tested successfully in several different observables and experiments, sometimes even at the per-mil level of precision. The latest results from the LEP2 experiment in 2005 showed however a quite sizable deviation ($\sim5\%$) from universality in the $W\ell\nu_\ell$ coupling of more than two sigmas when comparing the third leptonic family with the two light ones, as shown in Eq.~\eqref{eq:univvio}.
\par
We have considered in this article the possibility that this deviation represents a real NP effect. We have performed an Effective Field Theory analysis where the NP effects are parameterized by a series of Wilson coefficients $\alpha_i$, that appear in the effective Lagrangian multiplying dimension-six operators. In order to reduce the number of unknown coefficients and motivated by the possible deviation from lepton universality in the $W\tau\nu_\tau$ vertex, we have assumed different flavor symmetries where the third family plays a special role.
\par
Within this framework we have analyzed if it is possible to accommodate the $R^W_{\tau\ell}$ anomaly of Eq.~\eqref{eq:univvio} as a real NP effect without spoiling the nice agreement between SM predictions and EWPO observables. As expected, it is not possible to do such a thing with just one effective operator at play, due mainly to the very precise $Z$ and $\tau$ leptonic decays, as nicely shown in Fig.~\ref{fig:SingleOperator}. More surprisingly we have found that EWPO are such  strong constraints that not even in a global analysis where all the operators affecting the third family are present one can accommodate the $R^W_{\tau\ell}$ anomaly.
\par
Should this departure from universality be confirmed by new data, then our analysis disfavor the possibility of explaining it through a weakly coupled theory standing at the TeV scale, unless a quite non-trivial flavor structure occurs. Instead, it would be necessary to resort to a different description of NP that could involve the introduction of new light degrees of freedom or a strongly interacting sector with flavor dependent couplings to leptons. For example previous studies of this deviation from universality in $W$ decays have focused on the possibility that pair production of supersymmetric light charged Higgs bosons, almost degenerate with the $W$ and decaying largely into heavy fermions, could mimic $W \rightarrow \tau \, \overline{\nu}_{\tau}$ decays  \cite{Dermisek:2008dq,Park:2006gk}. Modifications on the electroweak gauge group in order to singularize the third family have also been considered \cite{Li:2005dc}.
\par
Last but not least we have shown the importance of the current measurements in leptonic and semileptonic $\tau$ decays as New Physics constraining observables that probe new directions in the parameter space of our EFT framework, and we have analyzed the sensitivity of the leptonic decays of pseudoscalar mesons to the the violations of lepton universality.

\subsection*{Acknowledgements}
The authors wish to thank Z.~Han for his help in the use of the Mathematica code provided by Ref.~\cite{Han:2005pr}, and T.~Dorigo, A.~Pich and M.~Schmitt and for useful comments and discussions. A.~F. and J.~P. are partially supported by MEC (Spain) under grant FPA2007-60323, by MICINN (Spain) under grant FPA2011-23778, by the Spanish Consolider-Ingenio 2010 Programme CPAN (CSD2007-00042) and by CSIC under grant PII-200750I026. A.~F. is also partially supported by a FPU grant by MEC (Spain). J.~P. is also partially supported by Generalitat Valenciana under grant PROMETEO/2008/069. MGA is supported by the U.S. DOE contract DE-FG02-08ER41531 and by the Wisconsin Alumni Research Foundation.

\appendix
\renewcommand{\theequation}{\Alph{section}.\arabic{equation}}
\renewcommand{\thetable}{\Alph{section}.\arabic{table}}
\setcounter{equation}{0}
\setcounter{table}{0}

\section{Theoretical expressions}
\label{app:A}
We collect here several theoretical results of observables that have been employed in our analyses and that we do not include in the main text.

\subsection{Leptonic $Z$ decays}
\label{app:b}
The effective Lagrangian contributing to the $Z\to\tau^+\tau^-$ decay in the flavour $U(2)^5$ symmetry group is given by:
\begin{eqnarray} \label{eq:leftzll}
 {\cal L}_{\mbox{\tiny EFT}}
 &=& i \, \overline{L}  D \! \! \! \! / L \,
 + \frac{1}{\Lambda^2} \left\{
    \widehat{\alpha}_{hL}^{1} {\cal O}_{hL}^{1}  + \widehat{\alpha}_{hL}^{3} {\cal O}_{hL}^{3}
 \right. \nonumber\\
 &&\hspace{-1.0cm}  \left.
 +   \widehat{\alpha}_{h \tau} {\cal O}_{h\tau} +  \widehat{\alpha}_{\tau W}^t {\cal O}_{\tau W}^t  + \widehat{\alpha}_{\tau B}^t {\cal O}_{\tau B}^t  + \mbox{h.c.} \right\} ,
\end{eqnarray}
and the corresponding decay width is:
\begin{eqnarray}\label{eq:widthzll}
\Gamma \left( Z \rightarrow \tau^+ \, \tau^- \right) & = & \frac{G_F   M_Z^3}{24 \, \sqrt{2} \, \pi}  \sqrt{1-4 \, z_{\tau}^2}  \nonumber\\
&&\hspace{-2.5cm} \times \left\{ \left( 1 - 4 \, v^- \right)
\left(1-4 \, z_{\tau}^2 \right) - 24 \, \sqrt{2} \, z_{\tau} \, t^Z \left( 4 s_W^2 - 1 \right)  \right. \nonumber \\
&& \hspace{-2.5cm}\biggl.  + \left[ \left( 4 s_W^2 - 1 \right) - 4 v^+ \right] \left( 4 s_W^2 - 1 \right) \left( 1+ 2 \,  z_{\tau}^2 \right) \biggr\} ,
\end{eqnarray}
where:
\begin{eqnarray}\label{eq:betas}
v^{-} &=& \alpha_{h \tau} - \alpha_{h L}^1 - \alpha_{hL}^3 , \nonumber \\
v^{+} &=& \alpha_{h \tau} + \alpha_{h L}^1 + \alpha_{hL}^3 , \nonumber  \\
t^{Z} &=& c_W  \alpha_{\tau W}^t + s_W  \alpha_{\tau B}^t  ,
\end{eqnarray}
being $s_W$ and $c_W$ the sine and cosine of the weak angle $\theta_W$ respectively, and $z_\tau=m_\tau\slash M_Z$.  As discussed in relation with Eq.~(\ref{eq:gammawlnu}) it can be noticed that the linear contribution of the tensor operators ${\cal O}_{\tau W}^t$ and ${\cal O}_{\tau B}^t$ is suppressed by the lepton mass over the Z mass. Moreover operators ${\cal O}_{h \tau}$, ${\cal O}_{hL}^1$ and ${\cal O}_{hL}^3$ simply modify the weight of the SM vertices.

\subsection{Anomalous magnetic moment of the tau lepton}
\label{app:c}
The tensor operators ${\cal O}_{\tau W}^{t}$ and ${\cal O}_{\tau B}^{t}$ provide a local contribution to the anomalous magnetic moment of the tau lepton.
The generic $\gamma \, \tau \, \overline{\tau}$ vertex is given by $- i \, e \, \varepsilon^{\mu}(q) \, \overline{u}(p') \, V_{\mu} \,
u(p)$ where:
\begin{eqnarray} \label{eq:vmugll}
 V_{\mu}
 &=&  F_1(q^2) \, \gamma_{\mu} \, + \, i \,  F_2(q^2) \,  \sigma_{\mu \nu} \, \frac{q^{\nu}}{2 \, m_\tau}\nonumber\\
 &  & ~+~ F_3(q^2) \, \gamma^5 \, \sigma_{\mu \nu} \, \frac{q^{\nu}}{2\,  m_\tau}~,
\end{eqnarray}
and the anomalous magnetic moment of the $\tau$ lepton is given by $a_\tau = \left( g_\tau - 2 \right) / 2 = F_2(0)$. By using
${\cal L}_{\mbox{\tiny EFT}}$ in Eq.~(\ref{eq:left}) we find the following expression for the $\tau$ lepton anomalous magnetic moment:
\begin{equation} \label{eq:al}
 a_\tau \, = \, a_\tau^{\mbox{\tiny SM}} \, + \, \frac{ 2 \, \sqrt{2}}{s_W} \, \frac{m_\tau}{M_W} \, \left( \, c_W \alpha_{\tau B}^t - s_W \alpha_{\tau W}^t \right) \, ,
\end{equation}
where the first term in the right-hand side is the SM contribution: $a_{\tau}^{\mbox{\tiny SM}} = 1.17721 (5) \times 10^{-3}$ \cite{Eidelman:2007sb}. The current experimental result is given by $-0.052 < a_{\tau} < 0.013$ at $95\%~\mbox{C.L.}$ \cite{Abdallah:2003xd}, though other analyses establish more stringent limits \cite{GonzalezSprinberg:2000mk}.


\end{document}